\RequirePackage{fix-cm}
\documentclass[smallextended]{svjour3}       
\smartqed  
\usepackage{graphicx}
\journalname{Experimental Astronomy}

\newcommand{\aap}{A\&A, }

\begin{document}

\title{The role of the background in past and future X-ray missions
}


\author{Silvano Molendi        
}


\institute{S. Molendi \at
              INAF/IASF-Milano \\
              Via Bassini 15, I-20133 Milano ITALY \\
              \email{silvano@iasf-milano.inaf.it}           
}

\date{Received: date / Accepted: date}

\maketitle

\begin{abstract}
Background has played an important role in X-ray missions, limiting the exploitation of
science data in several and sometimes unexpected ways. In this presentation I review
past X-ray missions focusing on some important lessons we can learn from them.
I then go on discussing prospects for overcoming
background related limitations in future ones.
\keywords{X-ray astrophysics \and Instrumentation:background \and Particle background \and Radiation environment \and Soft proton background}
\end{abstract}

\section{Introduction}
\label{intro}
Background has been a concern in X-ray astronomy right from the start. Indeed, during the rocket flight affording
the first detection of extra-solar X-rays \cite{Giacc:1962} three sources of counts were detected. A point source, later identified as Sco X-1, a diffuse emission coming from the whole sky, later identified as the unresolved emission coming mostly from AGN and, of
course, the instrumental background.

\section{Collimators}
\label{sec:col}
The first X-ray experiments to be flown on satellites, such as those on Uhuru, OSO-8 or HEAO1, used collimators
to achieve some degree of angular resolution. With such an approach, sensitivity could only be improved by increasing the collecting area of the instruments. Of course a large collecting area necessarily
implied a large instrumental background. Thus, the more advanced experiments of this kind were carefully designed to
to minimize background intensity and maximize background reproducibility.
The PDS experiment \cite{Frontera:1997}  on board BeppoSAX \cite{Boella:1997} is a good example of an advanced collimated detector. It operated in the hard X-ray band (15-200 keV) and comprised 4 PHOSWICH detectors plus a set of  anti-coincidence detectors. The angular resolution (FWHM of 1.3$^o$) was achieved through two rocking collimators, one for each pair of PHOSWICH detectors. During a typical observation, collimator A would point towards the source and collimator B would be offset towards background field 1. After a short time interval, collimator A would move to a background field 2  and B would go on source. After another time interval, collimator A would move back to on source and B would go to background field 1 and so on. This strategy allowed for a continuous monitoring of the background as well as control of systematics associated to the single collimator and background field. It is worth noting that such a strategy implied giving up half of the collecting area, that could be used on source, to the background monitoring. This is a pretty drastic choice.
The reason for it is that, as the designers of the PDS correctly anticipated, for the bulk
of the sources that would be observed by their instrument, the systematic errors on the background subtraction process would result in larger uncertainties than the statistical errors. Thus, doubling the collecting area would only lead to a better statistical characterization
of an already systematics limited observation, something not particularly useful and, as I can state from first hand experience,
very frustrating.
As it turns out the strategy adopted by the PDS was quite successful, systematic errors on the background were typically
kept below the percent level thereby allowing an unprecedented  exploration of the spectral properties of  several extragalactic sources.

\section{Concentrators}
\label{sec:con}

The adoption of X-ray mirrors changed the picture considerably. Telescopes, by concentrating photons on a small
spot, increased significantly the source to background ratio for point sources.
Naturally, the better the angular resolution of the telescope, the smaller the spot and the higher the ratio.
However, we should not forget that this is true as long as the angular extent of the source is smaller than the PSF.
It is  no longer true once we start resolving the source. In light of the substantial advantage concentrators afford,
hardware teams working on focused experiments have dedicated significantly less attention to reduction and
reproducibility of the background than those working on collimated ones.

\begin{figure*}
  \resizebox{0.97\hsize}{!}{\includegraphics[angle=-90]{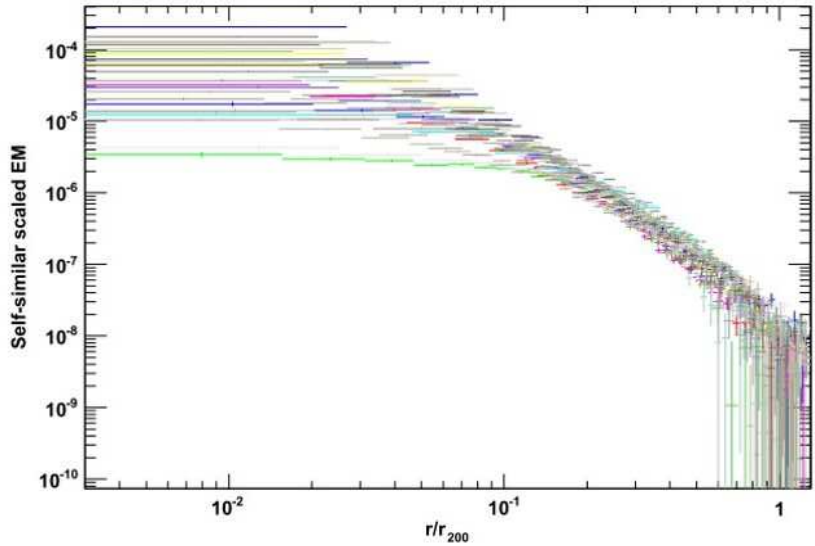}}
\caption{ Scaled emission measure (in units of cm$^{−6}$ Mpc) for a sample of 31 clusters, see \cite{Eckert:12} for more details.}
\label{fig:r}       
\end{figure*}

One exception to this rule is the PSPC experiment \cite{Pfeff:1987} onboard ROSAT. It was the first, and for 20 years only,
X-ray imaging experiment featuring an anticoincidence system. Moreover,
it was the only X-ray experiment where the instrumental background was subdominant with respect to the celestial background
over the full energy range. As a result of this, the ROSAT PSPC  sensitivity to low surface brightness emission has been unrivaled for over 25
years. An illustration of how important this is comes from the field of Galaxy Clusters. Some of the most accurate
measures of the Intra-Cluster Medium in Cluster outskirts, where emission is very tenuous, come from PSPC data
(see Fig.\ref{fig:r} and \cite{Eckert:12}).

While moving from collimators to concentrators reduced significantly the impact of the background,
moving from smaller to larger telescopes did not. This is because, thus far, efforts have concentrated
on increasing the effective area of X-ray telescopes, not their concentrating power. The concentrating power of a
telescope may be defined as the ratio of the effective area to the focal length. It should be noted
that increasing the concentrating power is far more difficult that increasing the effective area
for the simple reason that it is difficult to reflect X-rays at anything but very small angles.
However, it is also worth pointing out that very little effort has gone into increasing the concentrating
power. Indeed, multi-layer substrates, which allow to increase the maximum reflecting angle,
have thus far been adopted to extend imaging to the hard X-ray band not to increase the concentrating
power of medium energy experiments, possibly because the advantages this approach would entail
are poorly understood.

\section{Detector technology and Orbit}
\label{sec:det}

Over the last decades solid state detectors have substituted gas detectors, the major reason being
the significant improvement in spectral resolution. A gas detector like the GIS \cite{Makishima:1994} on ASCA
or the MECS \cite{Boella:1997} on BeppoSAX has a resolution of 8\% at 6 keV to be compared with a value of 2\% at 6 keV for
experiments like EPIC \cite{Turner:2001} on XMM or ACIS \cite{Garmire:1997} on Chandra. This is, of course, a substantial
improvement, but it does come at a price. Most solid state detectors that have been flown are much slower than gas detectors,
which implies that reduction of instrumental background through active shielding, is no longer an
option\footnote{Incidently this is one of the reasons why X-ray experiments such as ACIS on Chandra
or EPIC on XMM-Newton have a high background and limited sensitivity to low surface brightness emission.}.
Only in the last few years development of new technologies have led to solid state detectors
which, like those on NuSTAR \cite{Harrison:2013} and Hitomi \cite{Mitsuda:2014}, feature anti-coincidence systems.

Another factor affecting the background properties of detectors is the orbit.
High Earth Orbits, like those adopted for Chandra and XMM-Newton, have the advantage of
long uninterrupted observing windows however, as we shall see in the next section,
they also suffer from a higher and far less predictable background.
Conversely, Low Earth Orbits lead to a lower and more stable background, thanks to the shielding provided
by Earth's magnetic field.

\begin{figure*}
  \resizebox{0.97\hsize}{!}{\includegraphics[angle=-90]{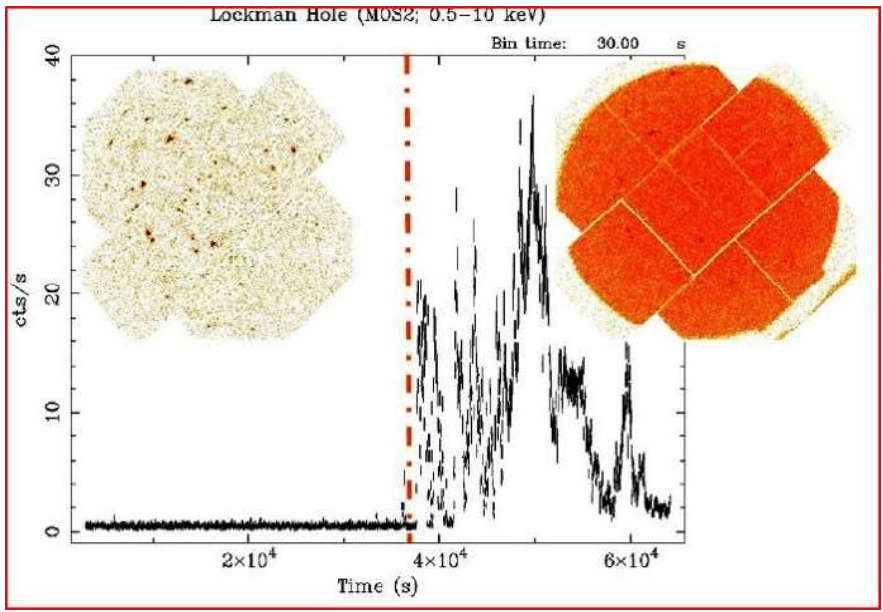}}
\caption{Main figure: lightcurve extracted from the full FoV of a MOS EPIC detector. Upper left inset: image
extracted from the first part of the observation. Upper right inset: image extracted from the second
part of the observation. Figure adapted from \cite{Tiengo:2007}.}
\label{fig:t1}       
\end{figure*}

\section{Soft Protons}
\label{sec:sp}

\begin{figure*}
  \resizebox{0.97\hsize}{!}{\includegraphics[angle=0]{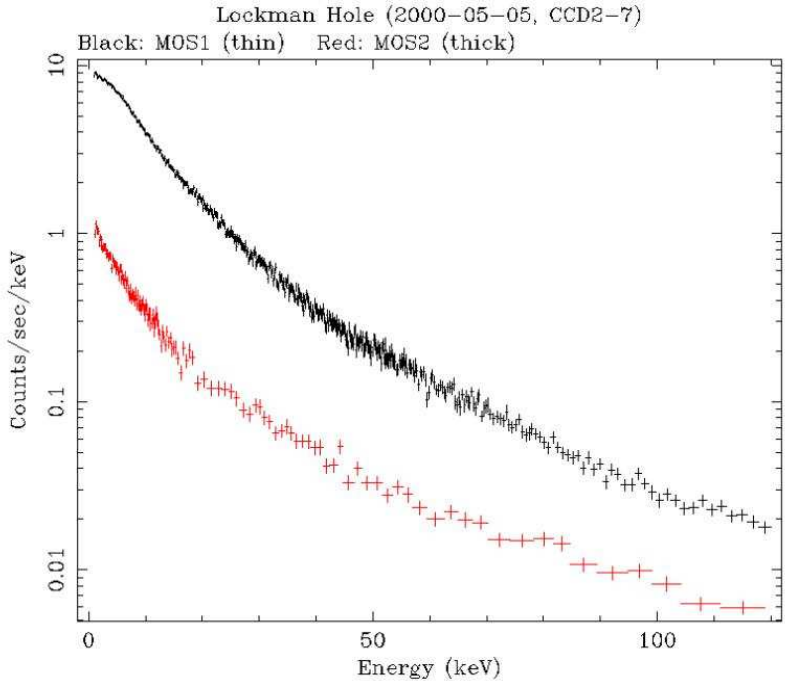}}
\caption{ MOS1 (black) and MOS2 (red) spectra accumulated in reduced gain mode during a soft proton flare.
See \cite{Tiengo:2007} for details.}
\label{fig:t2}       
\end{figure*}

In Fig.\ref{fig:t1} we show data collected in the early stages of the XMM-Newton mission.
The lightcurve, which is accumulated from the entire EPIC MOS field of view, shows a low
and stable rate up to about half of the observation. An image
produced during this period and reported in an inset on the upper left of the figure, shows
a typical EPIC survey field with several sources. During the second half of the observation
the rate shown in the lightcurve increases significantly showing variations of factors of 10 or more.
An image taken during this period, inset on the upper right
 of the image, shows a very different picture: the inFoV region is dominated by a diffuse emission
that  dwarfs emission from the sources. Scientists working on EPIC data were puzzled by
these events. Even more mindboggling was the fact that those ``flares" appeared to be totally unrelated
to the background rates measured by the radiation monitor onboard XMM-Newton.
Some light was finally shed when, as part of a series of  operational tests, the EPIC MOS detectors
were operated in reduced gain mode.  In this mode the EPIC detector can record events with  energies
up to $\sim$100 keV. Another key fact about this observation was the use of different filters
for the two MOS cameras. The thin filter for MOS1 and the thick filter for MOS2.
Andrea Tiengo noticed that the MOS1 and MOS2 spectra
accumulated during a period affected by strong flares, while similar in shape, appeared to be offset by a constant factor in energy,
roughly 30 keV (see Fig.\ref{fig:t2}). As it turned out, this is the difference in energy loss expected for protons with energies in the range of several tens to hundreds of keV as they go through the filters.  The particles were dubbed ”soft” protons to distinguish them from the much higher energy cosmic ray protons.
This component, which was unknown before the launch of Chandra and XMM-Newton\footnote{Actually this is not entirely correct, flares like those experienced by XMM-Newton had already been observed with the LE telescope onboard EXOSAT (e.g. \cite{Branduardi:1985}). Unfortunately no one seems to have paid any attention to them when the time came to design the XMM-Newton or Chandra
missions.}, represents an important addition to the background. Soft proton flares contaminate about 40\% of the
XMM-Newton observing time (for a detailed analysis see \cite{Salvetti_ahead:2017} in these proceedings).

\begin{figure*}
  \resizebox{0.97\hsize}{!}{\includegraphics[angle=0]{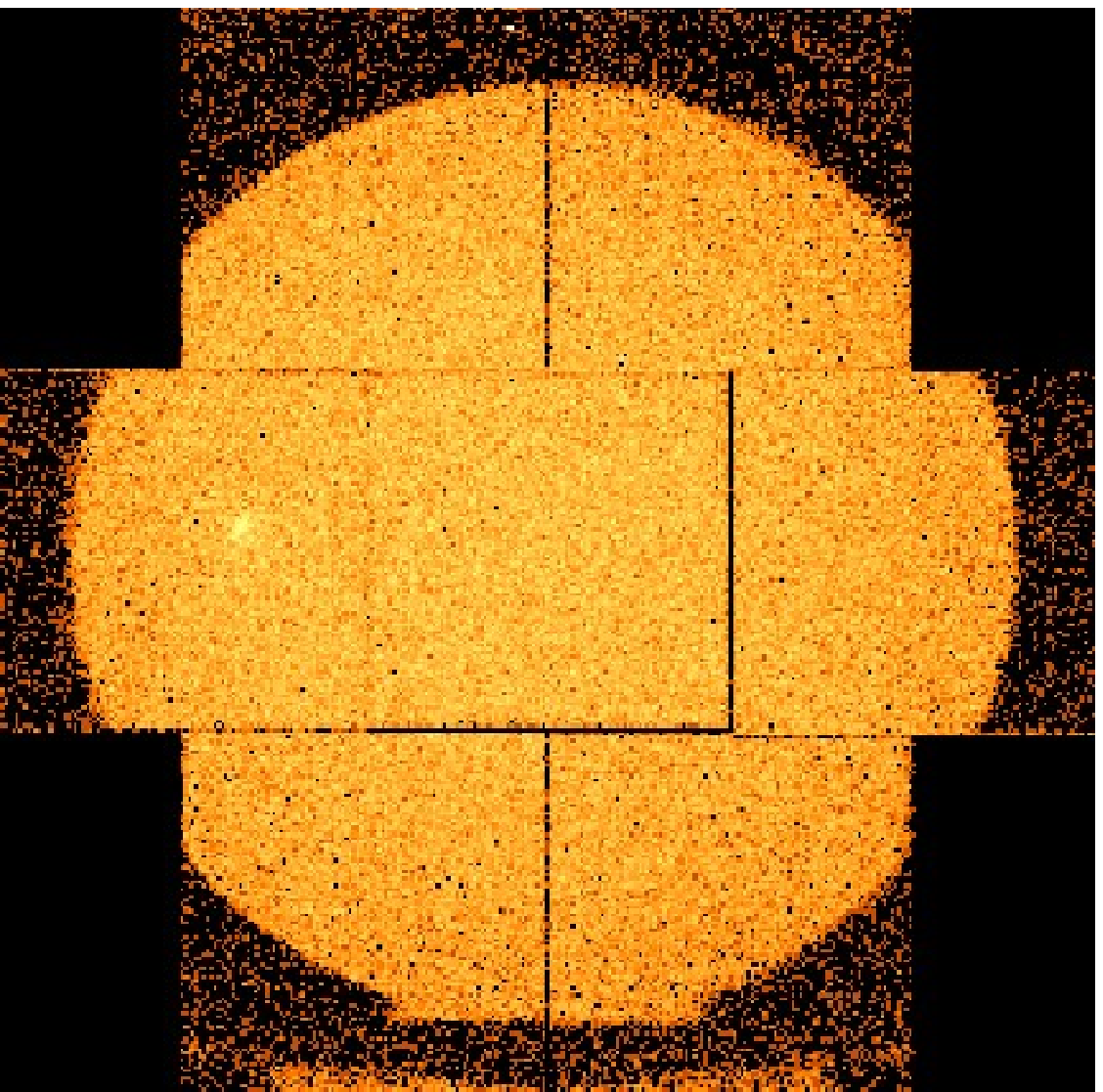}}
\caption{ MOS2 image accumulated in a hard energy band during a soft proton flare. Note the difference in surface brightness between
the inFoV region contaminated by soft protons and the outFoV where only high energy particle induced events
are recorded. }
\label{fig:dlm}       
\end{figure*}

One key feature of the EPIC MOS detectors is the presence of a significant fraction of detector area, about 30\%,
that is not exposed to the X-ray sky. By comparing the surface brightness inside the Field of View with that outside
we can separate the soft proton component, which contaminates only the inFoV region, from the high energy particle induced component, which is found both in the inFoV and the outFoV regions (see Fig.\ref{fig:dlm}).
In \cite{Deluca:2004} we found that, even during periods not contaminated by flares, the surface brightness inside the
Field of View  exceeded that recorded outside the Field of View by about 10\%.
It is worth noting that if the EPIC MOS detectors had not had an outFoV region this component would
have gone undetected.
Up to now it has been assumed that this low intensity component is related to soft protons, however recent work
\cite{Salvetti_ahead:2017} indicates otherwise.

\section{Future Missions: Athena}
\label{sec:fm}

\begin{figure*}
  \resizebox{0.97\hsize}{!}{\includegraphics[angle=0]{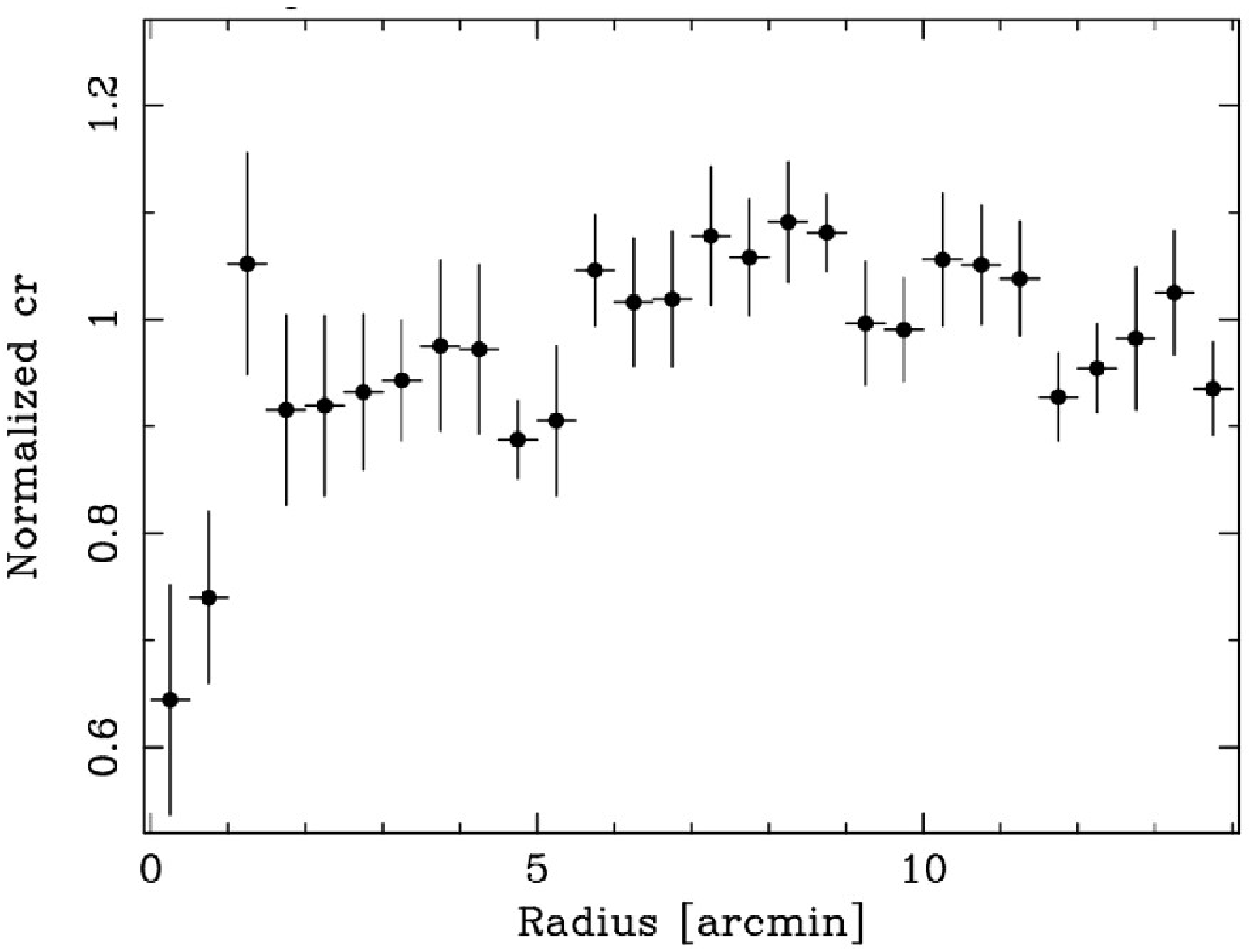}}
\caption{ Renormalized stacked radial surface brightness profile for a sample of 20 blank fields. Data is accumulated in the 0.7-1.2 keV band and from all 3 EPIC detectors, i.e. MOS1, MOS2 and pn. Given the random choice and large number of
blank fields employed we expect a flat profile.
With the exclusion of the innermost few arcminutes, where statistical handling of the data is less than optimal, the profile is always within a few percent of the expected value.}
\label{fig:sbe}       
\end{figure*}

Characterization of low surface brightness emission from the outer regions of Galaxy Clusters is
amongst the top science goals for ATHENA. Needless to say, work on the background, both the particle and the photon
component, will be vital for performing these kind of measures. This point is recognized by the ATHENA Science Study Team
(ASST), indeed one of the topical panels within the Mission Performance Working Group, one of five working
groups set up by the ASST, is dedicated to the study of background related issues.

\begin{figure*}
  \resizebox{0.97\hsize}{!}{\includegraphics[angle=90]{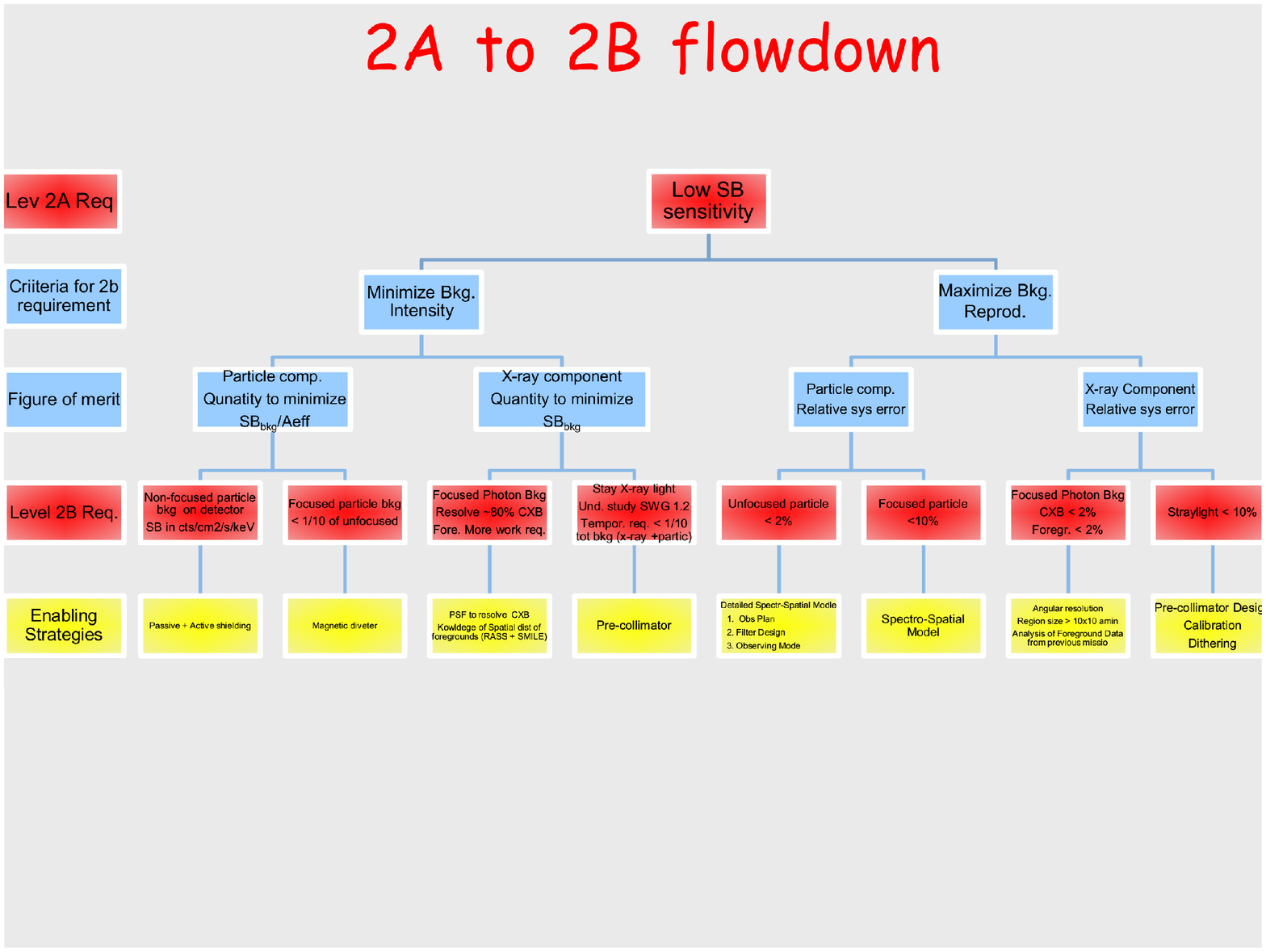}}
\caption{Flowdown from so called level 2a (mission independent) requirements to level 2b (mission dependent), see text
for details.}
\label{fig:flo}       
\end{figure*}

As should be apparent from what has been discussed in this proceeding, we can learn much about the background
that will affect experiments on ATHENA through the analysis of data accumulated during previous missions.
We have recently performed a systematic analysis of XMM-Newton EPIC archive within the framework of an ESA R/D activity
known as AREMBES (ATHENA Radiation Environment Models and X-Ray Background Effects Simulators). Within these
proceedings there are four contributions:\cite{Marelli_ahead:2017}, \cite{Salvetti_ahead:2017}, \cite{Ghizzardi_ahead:2017} and \cite{Gasta_ahead:2017}, dedicated to various aspects of this activity.

Work on the background should be organized around two criteria:
1) minimization of the background intensity; 2) maximization of background reproducibility.
Back of the envelope estimates and, later, detailed simulations (e.g. \cite{Molendi:2016}) indicate that the required level of reproducibility,
for the ATHENA instruments, is of a few percent. This is unprecedented, at least in as far as imaging experiments are concerned,
however, as we shall discuss in the next paragraphs, there is no reason why it cannot not be reached.
Indeed recent work that we have conducted on the EPIC background shows that even in the case of an experiment
that was not explicitly designed to maximize background reproducibility, careful analysis of the data can lead
to encouraging results (see Fig.\ref{fig:sbe} and \cite{Tchernin:2016}).

As already stated, background requirements for ATHENA
are challenging, however, they can be met by operating at 3 different levels: 1) experiment design; 2) observational strategy
and 3) data analysis strategy. At this time, points 1 and 2 must take precedence, however we need to maintain an overview of
the full process, a sort of holistic approach, if we wish to succeed. An attempt to provide this is reported in Fig.\ref{fig:flo}.
At the top of the figure we have a requirement on the sensitivity to surface brightness we wish to achieve. Below this we have the two criteria
previously discussed, i.e. minimization of intensity and maximization of reproducibility. On the next level,
we provide figures of merit, which differ for the different background components. Once these are defined we can
set specific requirements on different components. Finally, in the last row, we identify so called enabling
strategies that will allow to satisfy the requirements.

\section{Summary}
\label{sec:sm}

Whether we like it or not, background has been with us from the very start of X-ray astronomy.
Different missions have dealt with it in different ways and with varying degrees of success.
The examples I have provided indicate that a careful planing of the experiment is a key requisite
to keep background effects under control.
Attaining some of the top science goals for ATHENA will require a background reproducibility
which is unprecedented but not beyond reach.

\begin{acknowledgements}
The AHEAD project (grant agreement n. 654215) which is part of the EU-H2020 programm is
acknowledged for partial support.
\end{acknowledgements}

\bibliographystyle{spphys}       
\bibliography{biblio}   

%
%

\end{document}